\documentclass[aps,preprint,superscriptaddress,nofootinbib]{revtex4}
\usepackage{array}
\usepackage{graphicx}

\makeatother

\begin{document}
\newcommand{\fl}{\mathcal{F}_{L}}
 \newcommand{\fr}{\mathcal{F}_{R}}
 \newcommand{\g}{\mathcal{G}_{4f}}

\title{Search for New Physics via Single Top Production at the LHC}

\author{Qing-Hong Cao}

\email{qcao@ucr.edu}

\affiliation{Department of Physics and Astronomy, University of California at
Riverside, Riverside, CA 92521}

\author{Jose Wudka}

\email{jose.wudka@ucr.edu}

\affiliation{Department of Physics and Astronomy, University of California at
Riverside, Riverside, CA 92521}

\author{C.-P. Yuan}

\email{yuan@pa.msu.edu}

\affiliation{Department of Physics and Astronomy, Michigan State University, East
Lansing, MI 48824}

\begin{abstract}
We consider single-top production as a probe for new physics effects
at the Large Hadron Collider (LHC). We argue that for natural theories
a small deviation from the Standard Model tree-level couplings in
this reaction can be parameterized by 3 higher dimension operators.
Precision measurement of these effective couplings in the single-top
events, via studying their interference effects with the SM contributions,
can discriminate several new physics models. In particular, combining
the production rate of three single-top production modes will provide
a severe test of the Little Higgs model with T-parity. We find that
at the LHC, a $5\%$ accuracy in the measurement of the single-top
cross sections would probe the new physics scale up to about $3\,{\rm TeV}$. 
\end{abstract}
\maketitle
The search for New Physics (NP) beyond the Standard Model (SM) is
one of the major goals of the forthcoming Large Hadron Collider (LHC)
at CERN. The effects of new physics could be directly observed if
their characteristic scale lies below the center mass (CM) energy
of the relevant hard processes; otherwise they must be probed through
precision measurements of the SM couplings . When the available energy
is insufficient to directly produce the heavy excitations underlying
the SM, all new physics effects can be parameterized by the coefficients
of a series of gauge-invariant operators ($\mathcal{O}_{i}$) constructed
out of the SM fields~\citep{Georgi:1991ch,Weinberg:1978kz,Wudka:1994ny};
when the heavy physics decouples, as we will assume, these operators
have dimensions $\ge5$ and their coefficients are suppressed by inverse
powers of the new physics scale $\Lambda_{NP}$ (the scale at which
the excitations of the underlying theory can be directly probed)\ %
\footnote{The dimension 4 operators induced by the NP only renormalize the SM
coefficients and are unobservable, though relevant for naturalness
arguments. %
}.

The top quark, because of its heavy mass, is believed to provide a
good probe into new physics effects. In particular, processes containing
single top quark are expected to be sensitive to a rich variety of
physical effects. For instance, the corresponding production rates
can be significantly modified by NP interactions, such as heavy resonances
or non-standard flavor-changing vertices~\citep{Tait:2000sh}. In
the SM, single-top quark events can result from the $t$-channel process
($ub\to dt$), the $s$-channel process ($u\bar{d}\to t\bar{b}$)
and the $Wt$ associate production process ($bg\to tW^{-}$). Due
to their distinct kinematics, each of these three processes can be
differentiated and, in principle, measured separately. 
Recently, the evidence for single top quark production through weak
interactions has been reported by the D0 Collaboration at the Fermilab
Tevatron~\citep{Abazov:2006gd}. The soon-to-be-operational LHC offers
an excellent opportunity to search for NP via single top quark production.
The LHC will not only observe single-top events but also accurately
measure their characteristics. Since each single-top production process
will be affected differently by the NP effects, a comparison among
them can discriminate NP models.

In this letter we assume that NP effects in single-top production
will not be directly observed at the LHC (\textit{e.g.} as heavy resonances).
Such effects are then described by an effective Lagrangian of the
form \begin{equation}
\mathcal{L}_{eff}=\mathcal{L}_{SM}+\frac{1}{\Lambda_{NP}^{2}}\sum_{i}\left(c_{i}\mathcal{O}_{i}+h.c.\right)+O\left(\frac{1}{\Lambda_{NP}^{3}}\right),\label{eq:eft}\end{equation}
 where $c_{i}$'s are coefficients that parameterize the non-standard
interactions~%
\footnote{Dimension 5 operators involve fermion number violation and are assumed
to be associated with a very high energy scale and are not relevant
to the processes studied here.%
}. Because of the excellent agreement between the SM predictions and
precision experiments, the allowed deviations from the SM are small,
hence, when computing the effects of new operators we can restrict
ourselves to the interference terms between $\mathcal{L}_{SM}$ and
the operators $\mathcal{O}_{i}$, i.e. working to first order in the
coefficients $c_{i}$. Also, since the $c_{i}$ of loop-generated
operators are naturally suppressed by a numerical factor $\sim1/16\pi^{2}$,
we will only consider tree-level induced operators in this work.

\begin{figure}
\includegraphics[clip,scale=0.7]{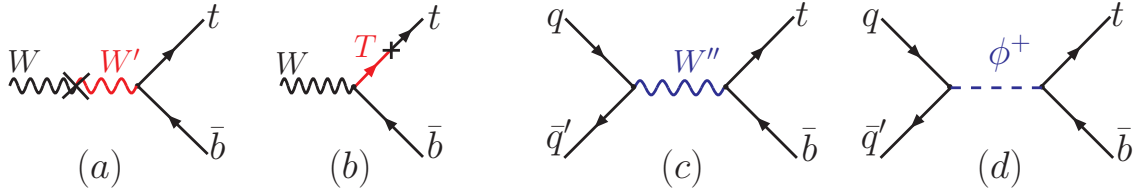} 

\caption{Examples of new physics that can induce the effective vertices listed
in Eqs.~(\ref{eq:fl-defin}) and (\ref{eq:operator-4f-2}). (a) and
(b) generate a $Wtb$ vertex through mixing with a heavy $W'$ gauge
boson or a heavy $T$ quark (top-quark partner), while (c) and (d)
induce effective four fermion operators through exchanging a heavy
$W'$ gauge boson or a heavy charged Higgs boson $\phi^{+}$. Although
(a)~and~(c) are both induced by $W'$, they originate from different
new physics effects: the former is related to the gauge boson mixing,
while the latter to the $W'$ couplings to quarks. \label{fig:illustration}}

\end{figure}

There are two types of tree-level induced effective operators that
contribute to single-top production: those modifying the $Wtb$ coupling,
which affect all production channels, and the four fermion interactions
that contribute only to the $s$-channel and $t$-channel production
processes; we will discuss them separately. For example, in Fig.~\ref{fig:illustration},
(a) and (b) modify the $Wtb$ vertex through mixing with a heavy $W'$
gauge boson or a heavy $T$ quark (top-quark partner), while (c) and
(d) induce effective four fermion operators through exchanging a heavy
$W'$ gauge boson or a heavy charged Higgs boson $\phi^{+}$.

As shown in Refs.~\citep{Buchmuller:1985jz,Arzt:1994gp,Cao:2006pu},
there are only 2 tree-level generated operators of the first type
that can contribute to single-top production: \begin{eqnarray}
\mathcal{O}_{\phi q}^{\left(3\right)} & = & i\left(\phi^{\dagger}\tau^{I}D_{\mu}\phi\right)\left(\bar{q}_{h}\gamma^{\mu}\tau^{I}q_{h}\right)+h.c.,\nonumber \\
\mathcal{O}_{\phi\phi} & = & i\left(\phi^{\dagger}\epsilon D_{\mu}\phi\right)\left(\bar{t}\gamma^{\mu}b\right)+h.c.,\label{eq:operatoer-wtb}\end{eqnarray}
 where $\phi$ denotes the SM scalar doublet, $D_{\mu}$ the covariant
derivative, $q_{h}$ the left-handed top-bottom $SU(2)$ doublet,
and $t(b)$ the corresponding right-handed isosinglets~\citep{Buchmuller:1985jz};
$\tau^{I}$ denote the usual Pauli matrices, and $\epsilon$ the two-dimensional
antisymmetric tensor ($\epsilon_{12}=-\epsilon_{21}=1$) in the weak
isospin space. Upon symmetry breaking, the above two operators generate
the following contribution to the $Wtb$ coupling: \begin{equation}
\mathcal{O}_{Wtb}=\frac{g}{\sqrt{2}}\left\{ \bar{t}\gamma^{\mu}\left(\fl P_{L}+\ \fr P_{R}\right)bW_{\mu}^{+}+h.c.\right\} ,\label{eq:fl-defin}\end{equation}
 where $\mathcal{F}_{L}=C_{\phi q}^{\left(3\right)}v^{2}/\Lambda_{{\rm NP}}^{2}$
and $\mathcal{F}_{R}=C_{\phi\phi}v^{2}/(2\Lambda_{{\rm NP}}^{2})$,
and $v=246\,{\rm {\rm GeV}}$ is the vacuum expectation value (VEV)
of $\phi$ .

There exists 3 tree-level-induced operators of the second type that
can contribute to single-top production ~\citep{Buchmuller:1985jz,Arzt:1994gp}:
\begin{eqnarray}
\mathcal{O}_{qu}^{(1)} & = & \left(\bar{q}_{l}t_{R}\right)\left(\bar{u}_{R}q_{l}\right),\label{eq:o4f1}\\
\mathcal{O}_{qq}^{(1)} & = & \left(\bar{q}_{l}^{i}t_{R}\right)\left(\bar{q}_{l}^{j}b_{R}\right)\epsilon_{ij},\label{eq:o4f2}\\
\mathcal{O}_{qq}^{(3)} & = & \frac{1}{2}\left(\bar{q}_{l}\gamma_{\mu}\tau^{I}q_{l}\right)\left(\bar{q}_{h}\gamma^{\mu}\tau^{I}q_{h}\right),\label{eq:operator-4f}\end{eqnarray}
where $q_{l}$ and $u_{R}$ denote either first or second generation
left-handed quark isodoublets and right-handed singlets, respectively.
The contributions of the first two of these operators, however, will
be of order of $c_{i}^{2}$ and can be ignored. This is because the
vertices generated by $\mathcal{O}_{qu}^{(1)}$ and $\mathcal{O}_{qq}^{(1)}$
do not interfer with the SM contribution when the bottom quark mass
is neglected. Hence we only need to consider the last operator, $\mathcal{O}_{qq}^{(3)}$,
from which we extract out the following effective $qq'bt$ vertex:
\begin{eqnarray}
\mathcal{O}_{4f} & = & \g\left[\frac{1}{v^{2}}\left(\bar{Q'}\gamma^{\mu}P_{L}Q\right)\left(\bar{b}\gamma_{\mu}P_{L}t\right)\right.\nonumber \\
 &  & \left.\,\,\,\,\,\,\,\,+\frac{1}{v^{2}}\left(\bar{Q}\gamma^{\mu}P_{L}Q'\right)\left(\bar{t}\gamma_{\mu}P_{L}b\right)\right],\label{eq:operator-4f-2}\end{eqnarray}
 where $\g=C_{qq}^{\left(3\right)}v^{2}/(2\Lambda_{{\rm NP}}^{2})$
and $Q,Q'$ denote light-flavor quarks ($u$, $d$, $c$, $s$). (We
have inserted $v^{2}$ to make $\g$ dimensionless.) For simplicity,
we assume that the coefficients of the four-fermion operators are
proportional to the SM Cabibbo-Koboyashi-Maskawa (CKM) matrix, i.e.
$C_{ud}^{(3)}=C_{cs}^{(3)}=kC_{us}^{(3)}=-kC_{cd}^{(3)}$\ with $k$
being equal to $1/\sin\theta_{c}$, where $\theta_{c}$ is the Cabibbo
angle %
\footnote{The numerical results presented below do not change noticeably when
$C_{us}^{(3)}=C_{cd}^{(3)}=0$.%
}.

It is important to note that the natural values for the coefficients
$\fl,$ $\fr$ and $\g$ is of order $\left(v/\Lambda_{NP}\right)^{2}$
and that the formalism is applicable whenever the CM energy for the
hard process, $\sqrt{\hat{s}}$, is significantly below $\Lambda_{NP}$.
Taking $\Lambda_{NP}\sim2\,{\rm TeV}$ we find the following estimates:
\begin{equation}
\left|\fl\right|,\left|\fr\right|,\left|\g\right|<0.01.\label{eq:coup_limits}\end{equation}
 Concerning the right-handed coupling in (\ref{eq:fl-defin}), it
is well known that recent data on the decay of $b\to s\gamma$ leads
to the constraint $\left|\fr\right|<0.004$~\citep{Chetyrkin:1996vx,Larios:1999au,Burdman:1999fw},
provided that other new-physics effects, such as those produced by
a $b\bar{s}t\bar{t}$ 4-fermion interaction~%
\footnote{This operator can be generated, for example, by exchanging a heavy
$W'$ vector boson.%
}, are absent. This constraint will still hold provided we assume (as
we will) that no cancellations occur between these two effects; in
this case all $\fr$ effects are negligible. Hence, we will restrict
ourselves to the effective vertices containing the couplings $\fl$
and $\g$ and examine their effects in various experimental observables.
In our calculation we will take all the effective couplings to be
real in order to simplify our analysis. We will also assume that the
$\nu\ell W$ vertex does not receive significant contributions from
physics beyond the SM. Finally, we note that in order to be consistent
with the LEP II experimental measurements of the asymmetry observables
$A_{FB}^{b}$ and $A_{LR}^{b}$~\citep{Yao:2006px}, the $W$$t_{L}$$b_{L}$,
$Z$$\bar{b}_{L}$$b_{L}$ and $Z$$\bar{t}_{L}$$t_{L}$ couplings
should be strongly correlated. The operator $\mathcal{O}_{\phi q}^{(3)}$,
of. Eq.~(\ref{eq:operatoer-wtb}), modifies the $W$$t_{L}$$b_{L}$
and $Z$$\bar{b}_{L}$$b_{L}$ couplings, at the same order of magnitude
as $\mathcal{F}_{L}$; however, the complete set of effective operators
includes $\mathcal{O}_{\phi q}^{\left(1\right)}=i\left(\phi^{\dagger}D_{\mu}\phi\right)\left(\bar{q}_{h}\gamma^{\mu}q_{h}\right)+h.c.$
(also tree-level induced \citep{Buchmuller:1985jz,Arzt:1994gp}),
which contributes to the $Z$$\bar{b}_{L}$$b_{L}$ and $Z$$\bar{t}_{L}$$t_{L}$
couplings. To agree with the LEP II data, the contributions from $\mathcal{O}_{\phi q}^{(1)}$
and $\mathcal{O}_{\phi q}^{(3)}$ to the $Z$$\bar{b}_{L}$$b_{L}$
coupling must cancel, in which case the $Z$$\bar{t}_{L}$$t_{L}$
coupling receives a modification of the same order as $\mathcal{F}_{L}$,
a prediction that can be tested at the LHC and future Linear Colliders
by measuring the associated production of $Z$ boson with top quark
pairs~\citep{Baur:2005wi}. In this paper we will not investigate
such effects.

\begin{figure}
\includegraphics[clip,scale=0.5]{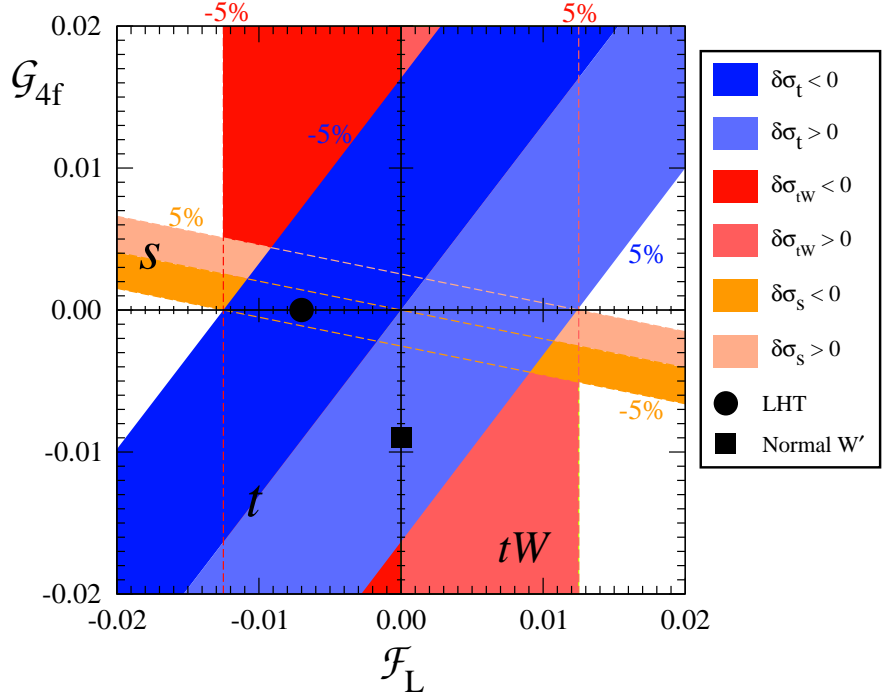}

\caption{Regions corresponding to $|\delta\sigma_{i}|\leq5\%$ for various
single-top production processes, in the plane of $\mathcal{F}_{L}$
and $\mathcal{G}_{4f}$. Predictions for two different models, LHT
(circle) and NP with heavy $W'$ (box), are also given. (See the main
text for its details.) \label{fig:3chan}}

\end{figure}

The explicit formulas for the inclusive cross sections of the three
single-top production channels at the LHC are found to be: \begin{eqnarray}
\sigma_{tW} & = & \sigma_{tW}^{0}\left(1+4\fl\right),\label{eq:sigma_tw}\\
\sigma_{s} & = & \sigma_{s}^{0}\left(1+4\fl+19.69\g\right),\label{eq:sigma_s}\\
\sigma_{t} & = & \sigma_{t}^{0}\left(1+4\fl-3.06\g\right),\label{eq:sigma_t}\end{eqnarray}
 while those for the Tevatron Run II are \begin{eqnarray}
\sigma_{tW} & = & \sigma_{tW}^{0}\left(1+4\fl\right),\label{eq:sigma_tw_tevatron}\\
\sigma_{s} & = & \sigma_{s}^{0}\left(1+4\fl+13.8\g\right),\label{eq:sigma_s_tevatron}\\
\sigma_{t} & = & \sigma_{t}^{0}\left(1+4\fl-2.2\g\right),\label{eq:sigma_t_tevatron}\end{eqnarray}
 where $\sigma_{i}^{0}$, with~$i=s,\, t,\, tW$ denote the SM cross
sections. The $\fl$ contribution is universal since it is associated
with a rescaling of the SM vertex. The four-fermion operators have
different effects in the $s$-channel and $t$-channel processes,
acting constructively or destructively (depending on the sign of $\mathcal{G}_{4f}$)
so that one process is always enhanced. The large coefficient in (\ref{eq:sigma_s})
indicates that the $s$-channel process is better suited for detecting
the effects of the operator containing $\mathcal{G}_{4f}$. 
The contribution of this operator in the top quark decay is negligible
because the SM contribution peaks in the region of phase space where
$(p_{\ell}+p_{\nu})^{2}\simeq M_{W}^{2}$, much smaller than $\Lambda_{NP}^{2}$.
As expected, the space-like $t$-channel exchange process is suppressed
by the large mass of the new particle, e.g. $Z^{\prime}$~\citep{Tait:2000sh,Sullivan:2002jt}.
The measurements can also determine the sign of $\g$. 
For illustration, we show in Fig.\ \ref{fig:3chan} the regions in
the$\mathcal{F}_{L}$- $\mathcal{G}_{4f}$ plane where the inclusive
cross sections of various single-top production processes deviate
from their corresponding SM cross sections, $\delta\sigma_{i}\equiv\left(\sigma_{i}-\sigma_{i}^{0}\right)/\sigma_{i}^{0}$,
by less than $5\%$ in magnitude, which we take this as a very rough
estimate of the systematic experimental uncertainty at the LHC\ \citep{Beneke:2000hk};
a realistic determination of this number must await the turning on
of the machine. It is worth noting that for the observables under
consideration, the NP effects can be comparable to the SM radiative
corrections, so we assume that all SM quantities are evaluated up
to the one-loop level, but the interference between the SM one-loop
and the new physics Born contributions can all be ignored.

Measuring each of the three production processes separately with sufficient
accuracy would allow for a complete determination of the $\fl$ and
$\g$ coefficients. In Table~\ref{tab:LHC-reach} we summarize the
LHC reach study of the single-top production by the ATLAS~\citep{SGTOP-LHC-LUCOTTE2}
and CMS~\citep{CMS-singletop-T,CMS-singletop-W} Collaborations.
Both studies clearly demonstrate that LHC has a great potential for
discovering all three single-top production processes and precisely
measuring their cross sections. In addition, we can derive the consistency
sum rule that the results must satisfy. It is \begin{equation}
\frac{\sigma_{s}}{\sigma_{s}^{0}}+6.43\frac{\sigma_{t}}{\sigma_{t}^{0}}=7.43\frac{\sigma_{tW}}{\sigma_{tW}^{0}}\,.\label{eq:sum-rule}\end{equation}
 In case of $\g=0$, Eq.~(\ref{eq:sum-rule}) becomes \begin{equation}
\frac{\sigma_{s}}{\sigma_{s}^{0}}=\frac{\sigma_{t}}{\sigma_{t}^{0}}=\frac{\sigma_{tW}}{\sigma_{tW}^{0}},\label{eq:sum-rule-lht}\end{equation}
 while in case of $\fl=0$, \begin{equation}
\frac{\sigma_{s}}{\sigma_{s}^{0}}+6.43\frac{\sigma_{t}}{\sigma_{t}^{0}}=0;\label{eq:sum-rule-wp}\end{equation}
 these relations can be used to discriminate new physics models, as
to be discussed below.

\begin{table}
\caption{Predicted event rates for various single-top production processes
by ATLAS and CMS Collaborations, where $S_{0}$ and $B$ denote the
numbers of the SM signal and background events, respectively. The
integrated luminosity ($\mathcal{L}$) is in the unit of ${\rm fb}^{-1}$.
$\frac{\sqrt{S_{0}+B}}{S_{0}}$ denotes the statistical uncertainty.
\label{tab:LHC-reach}}

\begin{tabular}{c|c|c|c|c|c|c|c}
\hline 
 &  & $S_{0}$ & $B$ & $\mathcal{L}$ & ${\displaystyle \frac{S_{0}}{B}}$ & ${\displaystyle \frac{S_{0}}{\sqrt{B}}}$ & ${\displaystyle \frac{\sqrt{S_{0}+B}}{S_{0}}}$\tabularnewline
\hline 
 & $t$ & 3130 & 925 & 10 & 3.38 & 325.4 & $2.0\%$\tabularnewline
ATLAS & $s$ & 385 & 2760 & 30 & 0.14 & 13.4 & $14.6\%$\tabularnewline
 & $Wt$ & 12852 & 133453 & 30 & 0.10 & 44.2 & $3.0\%$\tabularnewline
\hline 
 & $t$ & 2389 & 1785 & 10 & 1.34 & 179.8 & $2.7\%$\tabularnewline
CMS & $s$ & 273 & 2045 & 10 & 0.13 & 19.1 & $17.6\%$\tabularnewline
 & $Wt$ & 567 & 1596 & 10 & 0.36 & 44.9 & $9.2\%$\tabularnewline
\hline
\end{tabular}
\end{table}

For example, in the Little Higgs model with T-parity (LHT)~\citep{Cheng:2003ju,Cheng:2004yc,Low:2004xc},
the heavy gauge boson does not mix with the $W$-boson at tree-level,
so that $\fl$ can only be induced through the mixing of the top quark
with its even T-parity partner. In this theory $\Lambda_{NP}=4\pi f$
and, to first order in an expansion in powers of $v^{2}/f^{2}$, $\mathcal{F}_{L}=-c_{\lambda}^{4}v^{2}/(2f^{2})$
where $c_{\lambda}=\lambda_{1}/\sqrt{\lambda_{1}^{2}+\lambda_{2}^{2}}$
($\lambda_{1,2}$ denote the Yukawa couplings for the top quark and
its heavy partner); we also have $\mathcal{G}_{4f}=0$ so that (\ref{eq:sum-rule-lht})
can be used to restrict the other parameters. 
For example, taking $c_{\lambda}=1/\sqrt{2}$ and $f=1\,{\rm TeV}$,
yields $\mathcal{F}_{L}=-0.007$\ %
\footnote{We note that for this sample model of LHT, the predicted single-top
production rates for all three processes are smaller than the corresponding
SM rates. %
} and is represented by the circle in Fig.\ \ref{fig:3chan}. Hence,
the above analysis can be used to constrain the LHT parameters if
an excess in the single-top production rate is not found~\citep{Cao:2006wk}.

Another example is provided by the NP models that contain one or more
heavy, singly-charged vector-boson(s) $(W'$). Here we only consider
the simplest case where the $W'$ has the same couplings as the SM
$W$-boson. Recent Tevatron data on the search for $W'$ bosons in
the $t\bar{b}$ channel requires their mass be larger than $610\,{\rm {\rm GeV}}$~\citep{Abazov:2006aj}.
If we assume the $W'$ boson is much heavier and it does not mix with
the SM $W$ boson, the effective operator coefficients at the weak
scale will correspond to $\fl=0$ and $\g=-0.009$ when $\Lambda_{{\rm NP}}$
is taken to be $1200\,{\rm GeV}$. This model can be probed using
Eq.\ (\ref{eq:sum-rule-wp}), and is represented as the square in
Fig.\ \ref{fig:3chan}.

We will now argue that the statistical uncertainties in the measurement
of $\fl$ and $\g$ are quite small and the measurements will be dominated
by experimental uncertainties. To see this we temporarily ignore other
sources of uncertainty and follow the method described in\ \citep{Barger:2003rs}.
A reliable estimate of all errors would require a global analysis
of both the data and the properties of the detector using the same
philosophy as the one followed in Refs.~\citep{Pumplin:2001ct,Stump:2001gu}
for the analysis of the parton distribution functions. In this Letter,
however, our main purpose is to outline the methods for probing new
physics models via studying the single-top production rates. Hence
we will evaluate only the statistical uncertainties and simply assume
a $5\%$ experimental systematic uncertainty for all processes studied
here.  Needless to say that when data becomes available, a more comprehensive
analysis has to be carried out.

\begin{figure}[t]
 \includegraphics[clip,scale=0.6]{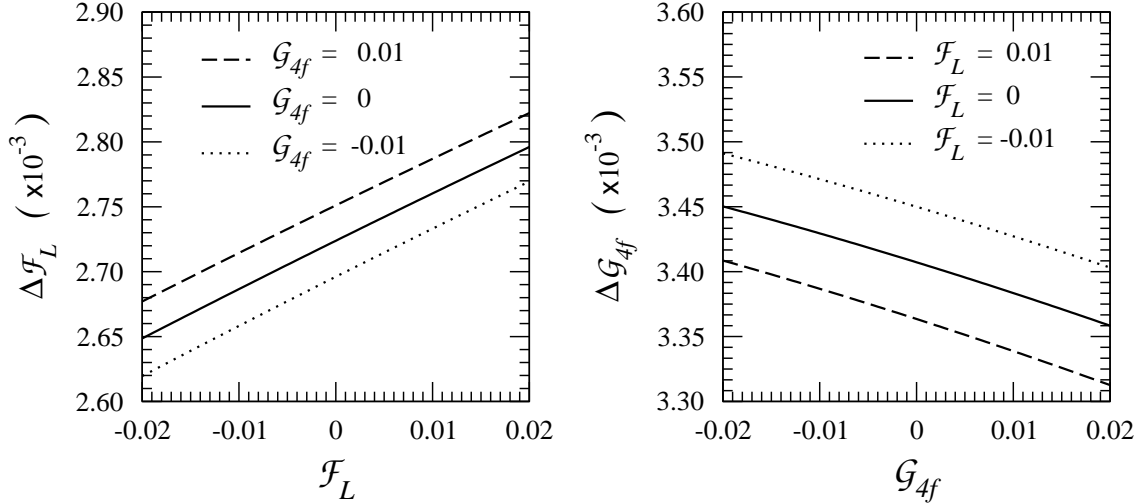} 

\caption{The expected statistical accuracy on measuring $\fl$ and $\g$ at
the ATLAS with an integrated luminosity of $30\,{\rm fb}^{-1}$ at
the LHC. \label{fig:observability}}

\end{figure}

It follows from (\ref{eq:sigma_tw}-\ref{eq:sigma_t}) that for each
single-top production channel the cross section can be expressed as
a product of the SM cross section, denoted as $\sigma^{0}$, and a
multiplicative factor depending linearly on the couplings $\fl$ and
$\g$: \begin{equation}
\sigma=\sigma^{0}\left(1+a\,\fl+b\,\g\right).\label{eq:linearEq}\end{equation}
We can then relate the accuracy of the cross section measurements
to the change of the effective couplings by \[
\frac{\Delta\sigma}{\sigma^{0}}=\left(a\,\Delta\fl+b\,\Delta\g\right),\]
 where ${\Delta\sigma}$ denotes the statistical uncertainty in the
measurement of $\sigma$, and $\Delta\fl$ and $\Delta\g$ denote
the corresponding quantities for $\fl$ and $\g$, respectively. Let
$S$ be the number of expected signal events for an integrated luminosity
$\mathcal{L}$ with $S=\sigma\mathcal{L}$, and $B$ the number of
background events (mainly from top quark pair production), we then
have \begin{eqnarray}
a\,\Delta\fl+b\,\Delta\g & \simeq & \frac{\sqrt{S_{0}+B}}{S_{0}}\left[1+\frac{S_{0}}{2(S_{0}+B)}\left(a\,\fl+b\,\g\right)\right]\\
 & \equiv & A.\label{eq:deltasigma2}\end{eqnarray}
 where $S_{0}=\mathcal{L}\sigma^{0}$. The last approximation holds
when $\left(a\fl+b\g\right)\ll1$ for all three single-top processes,
so that the limits on $\Delta\fl$, $\;\Delta\g$ will depend only
weakly on the values of $\mathcal{F}_{L}$ and $\mathcal{G}_{4f}$.
In this study, we consider one non-zero parameter at a time, so that
$\Delta\fl=A/a$ when $\Delta\g=0$, and $\Delta\g=A/\left|b\right|$
when $\Delta\fl=0$. The total statistical error after combining the
three channels in quadrature is \begin{equation}
\frac{1}{\Delta g}=\sqrt{\sum_{i=s,t,Wt}\frac{1}{\left(\Delta g_{i}\right)^{2}}},\label{eq:comb}\end{equation}
where $g$ denotes $\fl$ or $\g$. Due to their different experimental
setup, ATLAS and CMS have different sensitivities to the three channels.
In Fig.~\ref{fig:observability} we plot the statistical accuracy
on measuring $\fl$ and $\g$ at the ATLAS for $\mathcal{L}=30\,{\rm fb}^{-1}$.%
\footnote{ Here, we naively scale the signal and background event rates listed
in Table~I to those corresponding to $\mathcal{L}=30\,{\rm fb}^{-1}$
by the $\sqrt{\mathcal{L}}$ rule. %
} We find that this sensitivity can be quite high: for instance, if
$\fl=\g=0$ , $\Delta\fl\simeq0.0015$, which corresponds to a 0.2\%
accuracy in the measurement of the relevant SM couplings. As stated
above, these statistical uncertainties are much smaller than our rough
estimate of the experimental systematic errors. 

The sensitivity to each single-top production channel for $\fl=\g=0$
is presented in Table\ \ref{tab:uncertainty} for both the ATLAS
and CMS Collaborations. As explained above, the numerical results
will not change much for non-zero $\fl$ and $\g$. The $t$-channel
process provides the best measurement of $\fl$ both at ATLAS and
CMS in the sense that it has the smallest statistical uncertainty.
For the measurement of $\g$, contrary to the common belief, the reaches
of the $t$ and $s$-channel processes are comparable, because the
large coefficient of $\g$ in the $s$-channel process in Eq.~(\ref{eq:sigma_s})
compensates the larger uncertainty.

\begin{table}
\caption{The uncertainties $\Delta\fl$ and $\Delta\g$ for $\fl=\g=0$ with
$\mathcal{L}=30\,{\rm fb}^{-1}$. Here, only statistical uncertainty
is considered. \label{tab:uncertainty}}

\begin{tabular}{c||>{\centering}p{0.6in}|>{\centering}p{0.6in}|>{\centering}p{0.6in}|>{\centering}p{0.6in}}
\hline 
 & \multicolumn{2}{c|}{ATLAS} & \multicolumn{2}{c}{CMS}\tabularnewline
\hline 
 & $\Delta\fl$ & $\Delta\g$ & $\Delta\fl$ & $\Delta\g$\tabularnewline
\hline 
$t$-channel & 0.0029 & 0.0038 & 0.0039 & 0.0051\tabularnewline
$s$-channel & 0.0364 & 0.0074 & 0.0254 & 0.0052\tabularnewline
$tW$-channel & 0.0074 &  & 0.0118 & \tabularnewline
\hline
\end{tabular}
\end{table}

From the precision measurement of single-top events, one can also
derive conservative bounds on the new physics scales. Expressing the
deviations from the SM contributions, $\delta\sigma_{i}=\left(\sigma_{i}-\sigma_{i}^{0}\right)/\sigma_{i}^{0}$,
in terms of parameters which are more directly related to the heavy
physics, Eqs.~(\ref{eq:sigma_tw}-\ref{eq:sigma_t}) become \begin{eqnarray}
\delta\sigma_{tW} & = & 0.12C_{\phi q}^{(3)}\left(\frac{1{\rm TeV}}{\Lambda_{NP}}\right)^{2},\label{eq:bound-tw}\\
\delta\sigma_{s} & = & 0.12C_{\phi q}^{(3)}\left(\frac{1{\rm TeV}}{\Lambda_{NP}}\right)^{2}+0.60C_{qq}^{\left(1,3\right)}\left(\frac{1{\rm TeV}}{\Lambda_{NP}}\right)^{2},\qquad\label{eq:bound-s}\\
\delta\sigma_{t} & = & 0.12C_{\phi q}^{(3)}\left(\frac{1{\rm TeV}}{\Lambda_{NP}}\right)^{2}-0.09C_{qq}^{\left(1,3\right)}\left(\frac{1{\rm TeV}}{\Lambda_{NP}}\right)^{2}.\label{eq:bound-t}\end{eqnarray}
Though we expect $C_{i}=O(1)$, their precise values are unknown.
Measurements such as the ones described above can be used to obtain
the ratios of these coefficients, but the value of $\Lambda_{NP}$
cannot be obtained separately. After including the theoretical, statistical,
experimental systematic, and machine luminosity uncertainties, the
single-top processes are expected to be measured to a $5\%$ accuracy\ \citep{Beneke:2000hk}.
If we require $|\delta\sigma|\leq5\%$, then we obtain the following
realistic bounds \begin{equation}
\left|C_{\phi q}^{(3)}\right|\left(\frac{1{\rm TeV}}{\Lambda_{NP}}\right)^{2}<0.42,\quad\left|C_{qq}^{\left(3\right)}\right|\left(\frac{1{\rm TeV}}{\Lambda_{NP}}\right)^{2}<0.14\,,\label{eq:bound-result1}\end{equation}
 Assuming $C_{i}\simeq1$ these imply, \begin{equation}
\Lambda_{{\rm NP}}>2.8\,{\rm TeV}.\label{eq:bound-result2}\end{equation}
 It is worth noting that the average characteristic energy of the
hard processes is always significantly below $500\,{\rm GeV}$, for
the effective parton luminosity drops as the invariant mass of the
hard scattering process increases. Thus, the above results indicate
that single-top production provides a promising process which can
probe new physics effects up to $\sim6$ times the CM energy scale
of the hard scattering process.

\begin{figure}
\includegraphics[clip,scale=0.5]{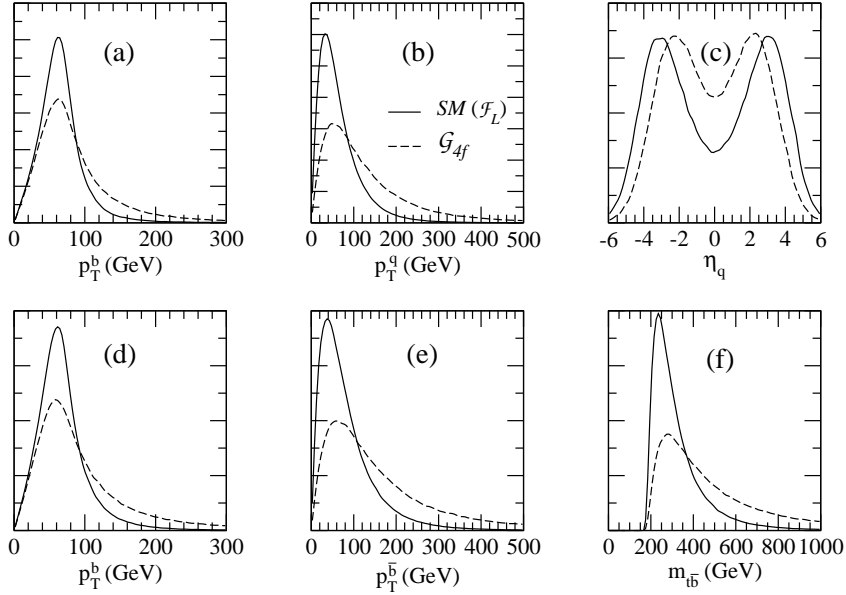} 

\caption{Normalized distributions of $p_{T}^{b}$, $p_{T}^{q}$ and $\eta_{q}$
of the $t$-channel process for $\g=-0.01$ (first row) and of $p_{T}^{b}$,
$p_{T}^{\bar{b}}$ and $m_{t\bar{b}}$ of the $s$-channel process
for $\g=0.01$ (second row) at the LHC. $p_{T}^{z}$ and $\eta_{z}$
denote the transverse momentum and rapidity of particle $z$; $m_{X}$
denotes the invariant mass of the set of particles $X$.\label{fig:lhc}}

\end{figure}

In the $t$-channel process the single-top quark is produced via the
$ub\to dt$ process with the subsequent decay of top quark $t\to bW^{+}(\to b\ell^{+}\nu)$.
Aside from the charged lepton and missing transverse energy, the final
state will contain two jets: one $b$-tagged jet and one non-$b$-tagged
light quark jet; the latter will be predominately in the forward direction
and can be used to suppress the copious SM backgrounds (such as those
produced by $Wb\bar{b}$ and $t\bar{t}$ events). In the $s$-channel
process, the single-top quark is produced via the $u\bar{d}\to t\bar{b}$
process with the subsequent top decay; its collider signature consists
of two $b$-tagged jets, one charged lepton, and missing transverse
energy. The transverse momentum ($p_{T}$) of the bottom quark from
top quark decay peaks at about $m_{{\rm t}}/3$ and it is insensitive
to the $\g$ coupling. In contrast, the $p_{T}$ distribution of the
$\bar{b}$ or $q$, produced in association with the $t$ quark is
shifted toward the large $p_{T}$ region by the $\g$ contribution;
a similar shift occurs in the invariant mass distributions of ($t$,$\bar{b}$)
system. The spectator jet is also shifted toward the central (for
$\g>0$) or forward (for $\g<0$) regions. These effects are illustrated
in Fig.~\ref{fig:lhc}.~%
\footnote{$\fl$ only produces a change in the overall normalization of the
cross section.%
}

The single-top production differential cross sections have been calculated
recently to NLO by various groups \citep{Harris:2002md,Sullivan:2004ie,Campbell:2004ch,Cao:2004ap,Cao:2004ky,Cao:2005pq,Frixione:2005vw,Campbell:2005bb,Beccaria:2006ir};
so the theoretical uncertainty in the SM predictions for the various
kinematical distributions is small. Extracting $\g$ from the corresponding
event distribution measurements will be limited mainly by experimental
statistics and systematic uncertainties, and is not expected to largely
improve the sensitivity obtained from the total cross-section measurements.

In summary, we have considered the single-top production at the LHC
as a probe for new physics effects. Assuming that the NP effects in
single-top production can not be directly observed as resonance enhancement
signal, we argued that for natural theories the small deviations from
the SM tree-level couplings in this reaction can be parameterized
by 3 couplings. One of these ($\fr$) is strongly constrained by the
low-energy data (assuming no cancellations), while another ($\fl$)
affects only the overall normalization of the single-top cross sections.
The four-fermion coupling $\g$ affects both the total cross section
and the kinematical distributions in the $s$- and $t$-channel processes,
acting constructively or destructively, depending on its sign. Accurate
measurement of all three production channels can determine $\fl$
and $\g$ to within a few percent (statistical) accuracy for an integrated
luminosity of $30\,{\rm fb}^{-1}$. The $s$-channel is expected to
be better suited for detecting $\g$ but suffers from larger statistical
and experimental uncertainties than the $t$-channel process. Our
study shows that the uncertainties of measuring $\g$ in the $s$-
and $t$-channel are comparable. Assuming the single-top production
can be measured with $5\%$ accuracy, one can probe the new physics
scale $\sim3\,{\rm TeV}$ in the single-top production at the LHC.

\textbf{Acknowledgments} We thank Ian Low for critical reading of
the manuscript and useful suggestions. We also thank Alexander Belyaev
and Ann Heinson for useful discussions. Q.-H. Cao and J. Wudka are
supported in part by the U.S. Department of Energy under grant No.
DE-FG03-94ER40837. C.-P. Yuan is supported in part by the U.S. National
Science Foundation under award PHY-0555545.

\bibliographystyle{apsrev} \bibliographystyle{apsrev}
\bibliography{reference}

\begin{thebibliography}{35}
\expandafter\ifx\csname natexlab\endcsname\relax\def\natexlab#1{#1}\fi
\expandafter\ifx\csname bibnamefont\endcsname\relax
  \def\bibnamefont#1{#1}\fi
\expandafter\ifx\csname bibfnamefont\endcsname\relax
  \def\bibfnamefont#1{#1}\fi
\expandafter\ifx\csname citenamefont\endcsname\relax
  \def\citenamefont#1{#1}\fi
\expandafter\ifx\csname url\endcsname\relax
  \def\url#1{\texttt{#1}}\fi
\expandafter\ifx\csname urlprefix\endcsname\relax\def\urlprefix{URL }\fi
\providecommand{\bibinfo}[2]{#2}
\providecommand{\eprint}[2][]{\url{#2}}

\bibitem[{\citenamefont{Georgi}(1991)}]{Georgi:1991ch}
\bibinfo{author}{\bibfnamefont{H.}~\bibnamefont{Georgi}},
  \bibinfo{journal}{Nucl. Phys.} \textbf{\bibinfo{volume}{B361}},
  \bibinfo{pages}{339} (\bibinfo{year}{1991}).

\bibitem[{\citenamefont{Weinberg}(1979)}]{Weinberg:1978kz}
\bibinfo{author}{\bibfnamefont{S.}~\bibnamefont{Weinberg}},
  \bibinfo{journal}{Physica} \textbf{\bibinfo{volume}{A96}},
  \bibinfo{pages}{327} (\bibinfo{year}{1979}).

\bibitem[{\citenamefont{Wudka}(1994)}]{Wudka:1994ny}
\bibinfo{author}{\bibfnamefont{J.}~\bibnamefont{Wudka}}, \bibinfo{journal}{Int.
  J. Mod. Phys.} \textbf{\bibinfo{volume}{A9}}, \bibinfo{pages}{2301}
  (\bibinfo{year}{1994}), \eprint{hep-ph/9406205}.

\bibitem[{\citenamefont{Tait and Yuan}(2001)}]{Tait:2000sh}
\bibinfo{author}{\bibfnamefont{T.}~\bibnamefont{Tait}} \bibnamefont{and}
  \bibinfo{author}{\bibfnamefont{C.-P.} \bibnamefont{Yuan}},
  \bibinfo{journal}{Phys. Rev.} \textbf{\bibinfo{volume}{D63}},
  \bibinfo{pages}{014018} (\bibinfo{year}{2001}), \eprint{hep-ph/0007298}.

\bibitem[{\citenamefont{Abazov et~al.}(2007)}]{Abazov:2006gd}
\bibinfo{author}{\bibfnamefont{V.~M.} \bibnamefont{Abazov}}
  \bibnamefont{et~al.} (\bibinfo{collaboration}{D0}), \bibinfo{journal}{Phys.
  Rev. Lett.} \textbf{\bibinfo{volume}{98}}, \bibinfo{pages}{181802}
  (\bibinfo{year}{2007}), \eprint{hep-ex/0612052}.

\bibitem[{\citenamefont{Buchmuller and Wyler}(1986)}]{Buchmuller:1985jz}
\bibinfo{author}{\bibfnamefont{W.}~\bibnamefont{Buchmuller}} \bibnamefont{and}
  \bibinfo{author}{\bibfnamefont{D.}~\bibnamefont{Wyler}},
  \bibinfo{journal}{Nucl. Phys.} \textbf{\bibinfo{volume}{B268}},
  \bibinfo{pages}{621} (\bibinfo{year}{1986}).

\bibitem[{\citenamefont{Arzt et~al.}(1995)\citenamefont{Arzt, Einhorn, and
  Wudka}}]{Arzt:1994gp}
\bibinfo{author}{\bibfnamefont{C.}~\bibnamefont{Arzt}},
  \bibinfo{author}{\bibfnamefont{M.~B.} \bibnamefont{Einhorn}},
  \bibnamefont{and} \bibinfo{author}{\bibfnamefont{J.}~\bibnamefont{Wudka}},
  \bibinfo{journal}{Nucl. Phys.} \textbf{\bibinfo{volume}{B433}},
  \bibinfo{pages}{41} (\bibinfo{year}{1995}), \eprint{hep-ph/9405214}.

\bibitem[{\citenamefont{Cao and Wudka}(2006)}]{Cao:2006pu}
\bibinfo{author}{\bibfnamefont{Q.-H.} \bibnamefont{Cao}} \bibnamefont{and}
  \bibinfo{author}{\bibfnamefont{J.}~\bibnamefont{Wudka}},
  \bibinfo{journal}{Phys. Rev.} \textbf{\bibinfo{volume}{D74}},
  \bibinfo{pages}{094015} (\bibinfo{year}{2006}), \eprint{hep-ph/0608331}.

\bibitem[{\citenamefont{Chetyrkin et~al.}(1997)\citenamefont{Chetyrkin, Misiak,
  and Munz}}]{Chetyrkin:1996vx}
\bibinfo{author}{\bibfnamefont{K.~G.} \bibnamefont{Chetyrkin}},
  \bibinfo{author}{\bibfnamefont{M.}~\bibnamefont{Misiak}}, \bibnamefont{and}
  \bibinfo{author}{\bibfnamefont{M.}~\bibnamefont{Munz}},
  \bibinfo{journal}{Phys. Lett.} \textbf{\bibinfo{volume}{B400}},
  \bibinfo{pages}{206} (\bibinfo{year}{1997}), \eprint{hep-ph/9612313}.

\bibitem[{\citenamefont{Larios et~al.}(1999)\citenamefont{Larios, Perez, and
  Yuan}}]{Larios:1999au}
\bibinfo{author}{\bibfnamefont{F.}~\bibnamefont{Larios}},
  \bibinfo{author}{\bibfnamefont{M.~A.} \bibnamefont{Perez}}, \bibnamefont{and}
  \bibinfo{author}{\bibfnamefont{C.-P.} \bibnamefont{Yuan}},
  \bibinfo{journal}{Phys. Lett.} \textbf{\bibinfo{volume}{B457}},
  \bibinfo{pages}{334} (\bibinfo{year}{1999}), \eprint{hep-ph/9903394}.

\bibitem[{\citenamefont{Burdman et~al.}(2000)\citenamefont{Burdman,
  Gonzalez-Garcia, and Novaes}}]{Burdman:1999fw}
\bibinfo{author}{\bibfnamefont{G.}~\bibnamefont{Burdman}},
  \bibinfo{author}{\bibfnamefont{M.~C.} \bibnamefont{Gonzalez-Garcia}},
  \bibnamefont{and} \bibinfo{author}{\bibfnamefont{S.~F.}
  \bibnamefont{Novaes}}, \bibinfo{journal}{Phys. Rev.}
  \textbf{\bibinfo{volume}{D61}}, \bibinfo{pages}{114016}
  (\bibinfo{year}{2000}), \eprint{hep-ph/9906329}.

\bibitem[{\citenamefont{Yao et~al.}(2006)}]{Yao:2006px}
\bibinfo{author}{\bibfnamefont{W.~M.} \bibnamefont{Yao}} \bibnamefont{et~al.}
  (\bibinfo{collaboration}{Particle Data Group}), \bibinfo{journal}{J. Phys.}
  \textbf{\bibinfo{volume}{G33}}, \bibinfo{pages}{1} (\bibinfo{year}{2006}).

\bibitem[{\citenamefont{Baur et~al.}(2006)\citenamefont{Baur, Juste, Rainwater,
  and Orr}}]{Baur:2005wi}
\bibinfo{author}{\bibfnamefont{U.}~\bibnamefont{Baur}},
  \bibinfo{author}{\bibfnamefont{A.}~\bibnamefont{Juste}},
  \bibinfo{author}{\bibfnamefont{D.}~\bibnamefont{Rainwater}},
  \bibnamefont{and} \bibinfo{author}{\bibfnamefont{L.~H.} \bibnamefont{Orr}},
  \bibinfo{journal}{Phys. Rev.} \textbf{\bibinfo{volume}{D73}},
  \bibinfo{pages}{034016} (\bibinfo{year}{2006}), \eprint{hep-ph/0512262}.

\bibitem[{\citenamefont{Sullivan}(2002)}]{Sullivan:2002jt}
\bibinfo{author}{\bibfnamefont{Z.}~\bibnamefont{Sullivan}},
  \bibinfo{journal}{Phys. Rev.} \textbf{\bibinfo{volume}{D66}},
  \bibinfo{pages}{075011} (\bibinfo{year}{2002}), \eprint{hep-ph/0207290}.

\bibitem[{\citenamefont{Beneke et~al.}(2000)}]{Beneke:2000hk}
\bibinfo{author}{\bibfnamefont{M.}~\bibnamefont{Beneke}} \bibnamefont{et~al.}
  (\bibinfo{year}{2000}), \eprint{hep-ph/0003033}.

\bibitem[{\citenamefont{Lucotte}(2006)}]{SGTOP-LHC-LUCOTTE2}
\bibinfo{author}{\bibfnamefont{A.}~\bibnamefont{Lucotte}},
  \bibinfo{journal}{ATLAS Note ATL-PHYS-COM-2006-000X} pp.
  \bibinfo{pages}{1--30} (\bibinfo{year}{2006}), \eprint{To appear in
  Proceedings of HCP 2005, TOP-2006}.

\bibitem[{\citenamefont{Abramov et~al.}(2006)\citenamefont{Abramov, Boos,
  Drozdetskiy, Dudko, Giammanco et~al.}}]{CMS-singletop-T}
\bibinfo{author}{\bibfnamefont{V.}~\bibnamefont{Abramov}},
  \bibinfo{author}{\bibfnamefont{E.}~\bibnamefont{Boos}},
  \bibinfo{author}{\bibfnamefont{A.}~\bibnamefont{Drozdetskiy}},
  \bibinfo{author}{\bibfnamefont{L.}~\bibnamefont{Dudko}},
  \bibinfo{author}{\bibfnamefont{A.}~\bibnamefont{Giammanco}},
  \bibnamefont{et~al.}, \bibinfo{journal}{CMS Note 2006/084}
  (\bibinfo{year}{2006}).

\bibitem[{\citenamefont{Blyth et~al.}(2006)\citenamefont{Blyth, Chao, Chen,
  Giammanco, Lei, Shu, and Yeh}}]{CMS-singletop-W}
\bibinfo{author}{\bibfnamefont{S.}~\bibnamefont{Blyth}},
  \bibinfo{author}{\bibfnamefont{Y.}~\bibnamefont{Chao}},
  \bibinfo{author}{\bibfnamefont{K.}~\bibnamefont{Chen}},
  \bibinfo{author}{\bibfnamefont{A.}~\bibnamefont{Giammanco}},
  \bibinfo{author}{\bibfnamefont{G.}~\bibnamefont{Lei},
  \bibfnamefont{Y.J.and~Petrucciani}}, \bibinfo{author}{\bibfnamefont{J.-G.}
  \bibnamefont{Shu}}, \bibnamefont{and}
  \bibinfo{author}{\bibfnamefont{P.}~\bibnamefont{Yeh}}, \bibinfo{journal}{CMS
  Note 2006/086}  (\bibinfo{year}{2006}).

\bibitem[{\citenamefont{Cheng and Low}(2003)}]{Cheng:2003ju}
\bibinfo{author}{\bibfnamefont{H.-C.} \bibnamefont{Cheng}} \bibnamefont{and}
  \bibinfo{author}{\bibfnamefont{I.}~\bibnamefont{Low}},
  \bibinfo{journal}{JHEP} \textbf{\bibinfo{volume}{09}}, \bibinfo{pages}{051}
  (\bibinfo{year}{2003}), \eprint{hep-ph/0308199}.

\bibitem[{\citenamefont{Cheng and Low}(2004)}]{Cheng:2004yc}
\bibinfo{author}{\bibfnamefont{H.-C.} \bibnamefont{Cheng}} \bibnamefont{and}
  \bibinfo{author}{\bibfnamefont{I.}~\bibnamefont{Low}},
  \bibinfo{journal}{JHEP} \textbf{\bibinfo{volume}{08}}, \bibinfo{pages}{061}
  (\bibinfo{year}{2004}), \eprint{hep-ph/0405243}.

\bibitem[{\citenamefont{Low}(2004)}]{Low:2004xc}
\bibinfo{author}{\bibfnamefont{I.}~\bibnamefont{Low}}, \bibinfo{journal}{JHEP}
  \textbf{\bibinfo{volume}{10}}, \bibinfo{pages}{067} (\bibinfo{year}{2004}),
  \eprint{hep-ph/0409025}.

\bibitem[{\citenamefont{Cao et~al.}(2006)\citenamefont{Cao, Li, and
  Yuan}}]{Cao:2006wk}
\bibinfo{author}{\bibfnamefont{Q.-H.} \bibnamefont{Cao}},
  \bibinfo{author}{\bibfnamefont{C.~S.} \bibnamefont{Li}}, \bibnamefont{and}
  \bibinfo{author}{\bibfnamefont{C.-P.} \bibnamefont{Yuan}}
  (\bibinfo{year}{2006}), \eprint{hep-ph/0612243}.

\bibitem[{\citenamefont{Abazov et~al.}(2006)}]{Abazov:2006aj}
\bibinfo{author}{\bibfnamefont{V.~M.} \bibnamefont{Abazov}}
  \bibnamefont{et~al.} (\bibinfo{collaboration}{D0}), \bibinfo{journal}{Phys.
  Lett.} \textbf{\bibinfo{volume}{B641}}, \bibinfo{pages}{423}
  (\bibinfo{year}{2006}), \eprint{hep-ex/0607102}.

\bibitem[{\citenamefont{Barger et~al.}(2003)\citenamefont{Barger, Han,
  Langacker, McElrath, and Zerwas}}]{Barger:2003rs}
\bibinfo{author}{\bibfnamefont{V.}~\bibnamefont{Barger}},
  \bibinfo{author}{\bibfnamefont{T.}~\bibnamefont{Han}},
  \bibinfo{author}{\bibfnamefont{P.}~\bibnamefont{Langacker}},
  \bibinfo{author}{\bibfnamefont{B.}~\bibnamefont{McElrath}}, \bibnamefont{and}
  \bibinfo{author}{\bibfnamefont{P.}~\bibnamefont{Zerwas}},
  \bibinfo{journal}{Phys. Rev.} \textbf{\bibinfo{volume}{D67}},
  \bibinfo{pages}{115001} (\bibinfo{year}{2003}), \eprint{hep-ph/0301097}.

\bibitem[{\citenamefont{Pumplin et~al.}(2002)}]{Pumplin:2001ct}
\bibinfo{author}{\bibfnamefont{J.}~\bibnamefont{Pumplin}} \bibnamefont{et~al.},
  \bibinfo{journal}{Phys. Rev.} \textbf{\bibinfo{volume}{D65}},
  \bibinfo{pages}{014013} (\bibinfo{year}{2002}), \eprint{hep-ph/0101032}.

\bibitem[{\citenamefont{Stump et~al.}(2002)}]{Stump:2001gu}
\bibinfo{author}{\bibfnamefont{D.}~\bibnamefont{Stump}} \bibnamefont{et~al.},
  \bibinfo{journal}{Phys. Rev.} \textbf{\bibinfo{volume}{D65}},
  \bibinfo{pages}{014012} (\bibinfo{year}{2002}), \eprint{hep-ph/0101051}.

\bibitem[{\citenamefont{Harris et~al.}(2002)\citenamefont{Harris, Laenen, Phaf,
  Sullivan, and Weinzierl}}]{Harris:2002md}
\bibinfo{author}{\bibfnamefont{B.~W.} \bibnamefont{Harris}},
  \bibinfo{author}{\bibfnamefont{E.}~\bibnamefont{Laenen}},
  \bibinfo{author}{\bibfnamefont{L.}~\bibnamefont{Phaf}},
  \bibinfo{author}{\bibfnamefont{Z.}~\bibnamefont{Sullivan}}, \bibnamefont{and}
  \bibinfo{author}{\bibfnamefont{S.}~\bibnamefont{Weinzierl}},
  \bibinfo{journal}{Phys. Rev.} \textbf{\bibinfo{volume}{D66}},
  \bibinfo{pages}{054024} (\bibinfo{year}{2002}), \eprint{hep-ph/0207055}.

\bibitem[{\citenamefont{Sullivan}(2004)}]{Sullivan:2004ie}
\bibinfo{author}{\bibfnamefont{Z.}~\bibnamefont{Sullivan}},
  \bibinfo{journal}{Phys. Rev.} \textbf{\bibinfo{volume}{D70}},
  \bibinfo{pages}{114012} (\bibinfo{year}{2004}), \eprint{hep-ph/0408049}.

\bibitem[{\citenamefont{Campbell et~al.}(2004)\citenamefont{Campbell, Ellis,
  and Tramontano}}]{Campbell:2004ch}
\bibinfo{author}{\bibfnamefont{J.}~\bibnamefont{Campbell}},
  \bibinfo{author}{\bibfnamefont{R.~K.} \bibnamefont{Ellis}}, \bibnamefont{and}
  \bibinfo{author}{\bibfnamefont{F.}~\bibnamefont{Tramontano}},
  \bibinfo{journal}{Phys. Rev.} \textbf{\bibinfo{volume}{D70}},
  \bibinfo{pages}{094012} (\bibinfo{year}{2004}), \eprint{hep-ph/0408158}.

\bibitem[{\citenamefont{Cao et~al.}(2005{\natexlab{a}})\citenamefont{Cao,
  Schwienhorst, and Yuan}}]{Cao:2004ap}
\bibinfo{author}{\bibfnamefont{Q.-H.} \bibnamefont{Cao}},
  \bibinfo{author}{\bibfnamefont{R.}~\bibnamefont{Schwienhorst}},
  \bibnamefont{and} \bibinfo{author}{\bibfnamefont{C.-P.} \bibnamefont{Yuan}},
  \bibinfo{journal}{Phys. Rev.} \textbf{\bibinfo{volume}{D71}},
  \bibinfo{pages}{054023} (\bibinfo{year}{2005}{\natexlab{a}}),
  \eprint{hep-ph/0409040}.

\bibitem[{\citenamefont{Cao and Yuan}(2005)}]{Cao:2004ky}
\bibinfo{author}{\bibfnamefont{Q.-H.} \bibnamefont{Cao}} \bibnamefont{and}
  \bibinfo{author}{\bibfnamefont{C.-P.} \bibnamefont{Yuan}},
  \bibinfo{journal}{Phys. Rev.} \textbf{\bibinfo{volume}{D71}},
  \bibinfo{pages}{054022} (\bibinfo{year}{2005}), \eprint{hep-ph/0408180}.

\bibitem[{\citenamefont{Cao et~al.}(2005{\natexlab{b}})\citenamefont{Cao,
  Schwienhorst, Benitez, Brock, and Yuan}}]{Cao:2005pq}
\bibinfo{author}{\bibfnamefont{Q.-H.} \bibnamefont{Cao}},
  \bibinfo{author}{\bibfnamefont{R.}~\bibnamefont{Schwienhorst}},
  \bibinfo{author}{\bibfnamefont{J.~A.} \bibnamefont{Benitez}},
  \bibinfo{author}{\bibfnamefont{R.}~\bibnamefont{Brock}}, \bibnamefont{and}
  \bibinfo{author}{\bibfnamefont{C.-P.} \bibnamefont{Yuan}},
  \bibinfo{journal}{Phys. Rev.} \textbf{\bibinfo{volume}{D72}},
  \bibinfo{pages}{094027} (\bibinfo{year}{2005}{\natexlab{b}}),
  \eprint{hep-ph/0504230}.

\bibitem[{\citenamefont{Frixione et~al.}(2006)\citenamefont{Frixione, Laenen,
  Motylinski, and Webber}}]{Frixione:2005vw}
\bibinfo{author}{\bibfnamefont{S.}~\bibnamefont{Frixione}},
  \bibinfo{author}{\bibfnamefont{E.}~\bibnamefont{Laenen}},
  \bibinfo{author}{\bibfnamefont{P.}~\bibnamefont{Motylinski}},
  \bibnamefont{and} \bibinfo{author}{\bibfnamefont{B.~R.}
  \bibnamefont{Webber}}, \bibinfo{journal}{JHEP} \textbf{\bibinfo{volume}{03}},
  \bibinfo{pages}{092} (\bibinfo{year}{2006}), \eprint{hep-ph/0512250}.

\bibitem[{\citenamefont{Campbell and Tramontano}(2005)}]{Campbell:2005bb}
\bibinfo{author}{\bibfnamefont{J.}~\bibnamefont{Campbell}} \bibnamefont{and}
  \bibinfo{author}{\bibfnamefont{F.}~\bibnamefont{Tramontano}},
  \bibinfo{journal}{Nucl. Phys.} \textbf{\bibinfo{volume}{B726}},
  \bibinfo{pages}{109} (\bibinfo{year}{2005}), \eprint{hep-ph/0506289}.

\bibitem[{\citenamefont{Beccaria et~al.}(2006)\citenamefont{Beccaria, Macorini,
  Renard, and Verzegnassi}}]{Beccaria:2006ir}
\bibinfo{author}{\bibfnamefont{M.}~\bibnamefont{Beccaria}},
  \bibinfo{author}{\bibfnamefont{G.}~\bibnamefont{Macorini}},
  \bibinfo{author}{\bibfnamefont{F.~M.} \bibnamefont{Renard}},
  \bibnamefont{and}
  \bibinfo{author}{\bibfnamefont{C.}~\bibnamefont{Verzegnassi}},
  \bibinfo{journal}{Phys. Rev.} \textbf{\bibinfo{volume}{D74}},
  \bibinfo{pages}{013008} (\bibinfo{year}{2006}), \eprint{hep-ph/0605108}.

\end{thebibliography}

\end{document}